# A simple beam combination for stellar interferometry

E. N. Ribak

*Physics Department, Technion - Israel Institute of Technology, Haifa 32000, Israel*

M. Gai, D. Loreggia

*Osservatorio Astronomico di Torino - INAF, Via Osservatorio, 20, 10025 Pino Torinese, Italy*

S. G. Lipson

*Physics Department, Technion - Israel Institute of Technology, Haifa 32000, Israel*

In stellar interferometry, image quality improves significantly with the inclusion of more telescopes and the use of phase closure. We demonstrate, using first coherent and then partially coherent white light, a compact and efficient pair-wise combination of twelve or more beams. The input beams are lined up and spread through a cylindrical lens into a comb of parallel ellipses, which interferes with a perpendicular copy of itself to form a matrix of interferograms between all pairs. The diagonal elements show interference of each beam with itself, for intensity calibration. The measured white-light visibilities were high and stable. © Optical Society of America

A problem common to all multi-aperture astronomical interferometers is that of efficiently combining the signals in a pair-wise manner, i.e. such that each signal interferes independently with each one of the others. If there are $N$ apertures, there will then be $N(N-1)/2$ output interference signals, each one providing information on the coherence function for the appropriate baseline, corresponding to one Fourier component of the observed object. The equivalent problem in radio astronomy has been more easily solved, because in that region the number of photons per coherent mode is large, and so the electromagnetic fields received at the apertures can be measured directly and cloned. Unfortunately, the paucity of photons in the optical and infra-red regions does not permit the same solution.

Present-day stellar interferometers combine the light waves from many apertures using either bulk or integrated optics[1-4]. Pair-wise (Michelson) combination of $N$ beams arriving from $N$ apertures requires at least $N\log_2(N-1)+N(N-1)/2$ beam-splitters, each of which introduces wavelength- and polarization-dependent losses. In some systems, pair-wise combination is not implemented, and the individual interference signals are separated by using spatial (Fizeau) or temporal coding[1-7], where the other beams contribute to the background. Photon losses and noise become considerable in all the above techniques when $N$ is large, and a more efficient approach, independent of $N$, is described in this letter.

The system we present here allows a large number of beams to interfere pair-wise by the use of two beam-splitters. We tested the idea first by using laser sources, and following that, with partially coherent white-light sources, a necessary requirement for astronomical application. We use an anamorphic approach in which a linear array of $N$ concentrated beams from the individual sub-apertures is stretched in the direction normal to its linear axis (Fig. 1) using cylindrical optics, giving an array of beams with long elliptical cross-section. After a first beam-splitting, the two copies of this array are rotated mutually by $90^0$ and interfere at a second beam-splitter. Thus every elongated beam intersects each other one, giving a square array of interference patterns. Each pattern has interference contrast and phase corresponding to the two aperture beams. The diagonal elements, in which each beam interferes with itself, have high contrast and can be used for instrumental and aperture intensity calibration.

The idea was demonstrated using a quadrilateral Sagnac common-path interferometer into which a non-planar beam rotator was introduced[8-10]. The advantages of this type of interferometer are that it is mechanically stable, and that because of the common path, white-light fringes are easy to obtain. In the first version of the experiment (Fig. 2a), the beam rotation was obtained by inserting a Dove prism[8,9], which rotated the counter-propagating beams by $\pm 45^0$. Thus the interferometric system involved two beam-splits and four additional reflections (one in the Dove prism). The interference pattern was sensed by a CCD camera.

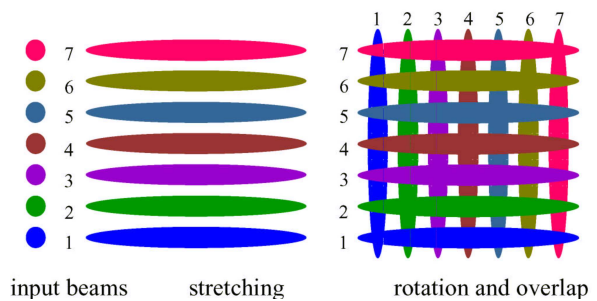

Fig. 1. (Color online) The telescopes beams are lined up (left), and stretched in the orthogonal direction to form a comb (center). This comb is then interfered with itself rotated at a normal angle. All fringes from all beams are measured on one detector (right).

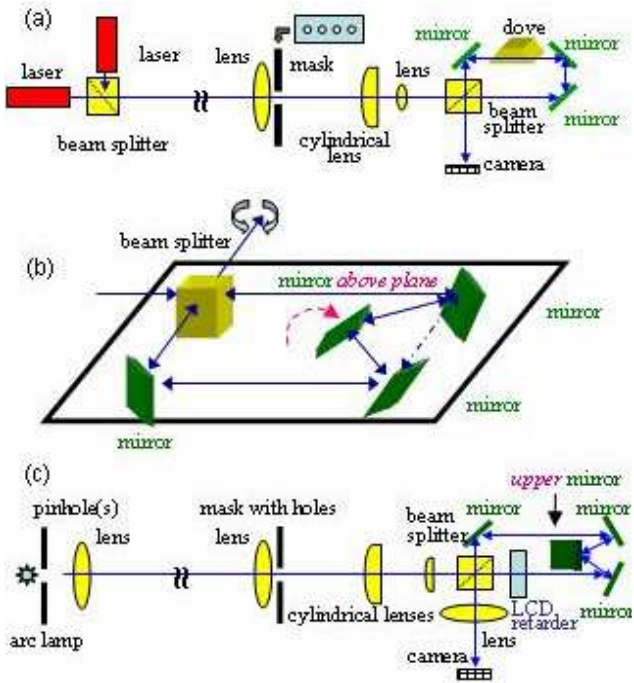

Fig. 2. (Color online) Laboratory setups: (a) A (laser) binary, a telescope, and a beam combiner: anamorphic optics and a Sagnac $90^0$ shear interferometer. (b) Prism-less rotational shear interferometer. One mirror is above the plane of the rest of the elements. (c) White-light source(s), including anamorphic telephoto and the new shear interferometer.

A linear set of six apertures was first illuminated by a plane wave from a laser, and high contrast fringes were observed at all intersections. In the interference pattern (Fig. 3, left) the contrast of the interference on the diagonal is high, while on the off-diagonal elements it is modulated symmetrically. The apertures were then illuminated by plane waves from two independent lasers, combined at a small angle by means of a non-polarizing beam-splitter (Fig. 2a), in order to simulate the coherence properties of waves from a binary star. The resulting contrast of the off-diagonal elements is modulated symmetrically and periodically, as expected (Fig. 3, right). Weak light leakage is evi-dent between the lines due to the inexpensive cylindrical optics employed.

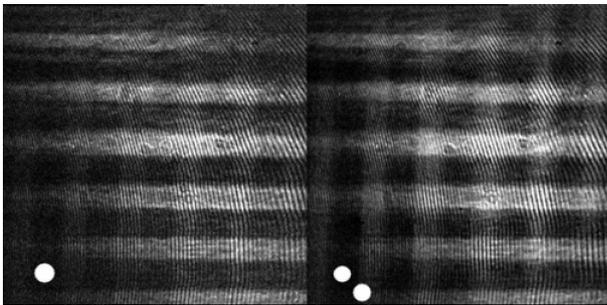

Fig. 3. The interference pattern of a single laser beam ("single star", left) and two lasers ("binary", right). High-contrast fringes are visible along the diagonal (bottom left to top right), where each beam interferes with itself. For the "single star" the contrast remains high but for the "binary" it periodically drops off and rises again with distance from the diagonal. The artificial fringe pattern is a product of the curved wave front alone. However, the phase of these fringes and their contrast change according to the source and intervening atmosphere.

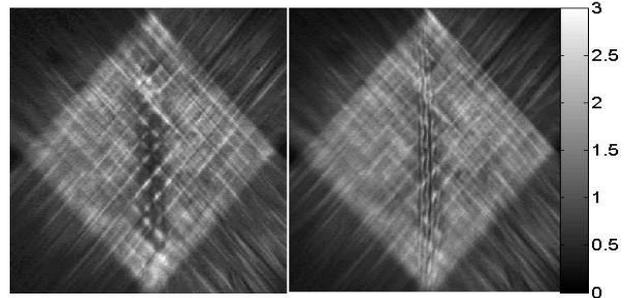

Fig. 4. The interference pattern of a single (left) and double (right) white light source. The diagonal is now vertical.

In a second experiment using white light, the interferometer was modified by using a fourth out-of plane mirror instead of the Dove prism[11] (Fig. 2b) and by dropping the fringe density. In this way, chromatic effects due to the prism refraction and to path difference were reduced. The white light arc source (Solarc) was imaged through one or two small apertures in order that the apertures at the entrance to the system would be illuminated partially coherently (Fig. 2c). To further condense these apertures we used a linear array of twelve very weak lenslets, which caused some curving of the fringes inside each beam (Fig. 4).

The visibility can be taken of the whole pattern, or within each pair intersection, by spatial or temporal modulation. However, we found out that by changing the polarization retardation within the Sagnac interferometer, the phases of the fringes vary. We obtained a series of images corresponding to different phase changes. We then calculated the average raw visibility, and ignored all pixels where the average intensity was negligible (Fig. 5). We also calculated the maximum visibility obtained at different phases (Fig. 6).

Previous experiments[8-12] and calculations[13] indicate a strong dependence of polarization on reflective layer material and out-of-plane geometry. A retarder inserted in the Sagnac-mirrors loop (Fig. 2c) was adjusted for better contrast. However, only one half of the light was utilized, and half sent back to the source. To also take advantage of the other (symmetric) half, we are looking into the possibility of a preceding polarizing beam splitter, where the two polarizations are sent into separate Sagnac combiners. Their four outputs can yield directly the phase of each fringe. Alternatively, the two asymmetric (lower contrast) outputs can serve to track the fringes mechanically, allowing longer integration of the two symmetric outputs for better signal-to-noise ratio (SNR).

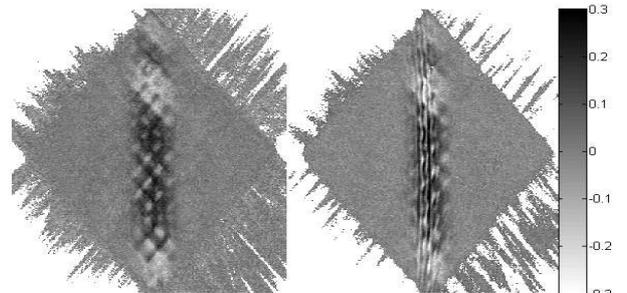

Fig. 5. The raw visibility of the fringes, corresponding to Fig. 4. The first lobe is barely visible (left), and the modulation due to the second source is clear (right).

The spaces between the spread beams reduce the total efficiency by $\sim g/(g+d)$, where $g$ is the gap and $d$ the beam diameter. The beams can be as close as the thickness of the mounting of the last lenses, or even adjacent for back-mounted mirrors. As more beams are added, they become more elongated and less light is lost between them. Only mirror size and scarcity of photons seem to limit the number of beams.

In pair-wise combination the light is divided between the outputs of the combiner and the SNR drops. The $N^2$ intersections are roughly twice the minimum required for all two-beam combinations[1-3, 9-11]. Polarization splitting of the beams into two interferometers recovers effectively most of the flux, which is then detected in four detectors. Redundancy is implicit in any combination scheme, because of the complementary outputs, but here the noise contribution is doubled again to four. Hence the noise equivalent magnitude is degraded by 1.5 for the faintest objects. For example, when the read out noise is ~25 photons, a signal with SNR = 10 on one pixel generates an equivalent SNR ~ 8 over four pixels; even lower readout noise produces even better SNR. In addition, integration time depends only on atmospheric (or multimode fiber) stability, as common or separate modulation of the beams is not required.

The degradation, due to the one added division, is mitigated by the high throughput of the proposed scheme. Assuming 98% reflection for two cylindrical and four Sagnac mirrors, 6% total loss on the beam-splitter/combiner, and 4% loss on the retarder, the total transmission budget is 80%, for any number of beams. With light loss in the gaps between the beams, this can drop to 63%. Traditionally combining only eight beams, each beam passes 15 beam-splitters and 14 mirrors, and half of the light is lost because of polarization, leading to a total efficiency of 22%.

We designed a new off-axis anamorphic telescope, with a significantly reduced field curvature and without vignetting. Two cylindrical off-axis parabolas were optimized in a central field: the beam is expanded in one dimension only, by two toroid surfaces, with no light spilling (Fig. 7). The root-mean-square phase error over the ellipse is 0.31 nm, and can be calibrated away. Finally, because of the near-normal reflection, polarization is conserved quite accurately.

In summary, the first experimental test of the aperture di-vision beam combiner was shown to be successful. With basic optical components we were able to produce stable, high-contrast fringes both on coherent independent sources and on white-light ones. A single camera was used to capture six to twelve beam combinations simultaneously. Future laboratory work will include beams arriving from a two-dimensional array of apertures, requiring their lining up into the interferometer. Finally, we will apply turbulence and analyse the result to obtain the full calibrated visibility, and hence phase closure relations.

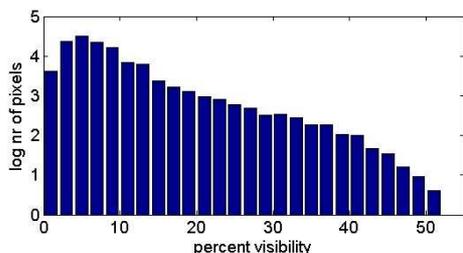

Fig. 6. Histogram of the maximum visibility at every pixel, over many fringe phases, of a single source (Fig. 4).

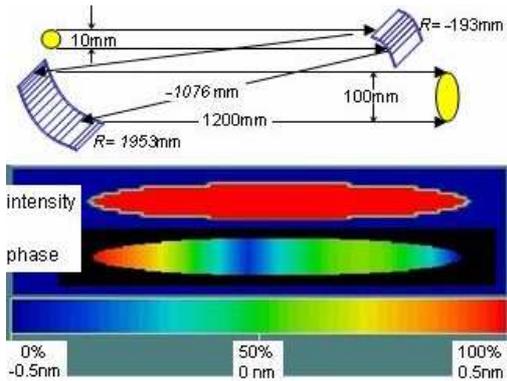

Fig. 7. (Color online) Anamorphic beam expander design with cylindrical off-axis parabolas, 50 mm apart (top, not to scale). The resulting intensity (center) is flat, as is the phase (bottom), to nanometer accuracy.

Acknowledgements. ER and MG both wish to thank the Fizeau Visitor Exchange Program of the European Interferometry Initiative and OPTICON (a European Framework VI programme).

**References**
1. M. Shao and M. M. Colavita, Ann. Rev. Astron. Astroph. **30**, 457 (1992).
2. J. D. Monnier, Rep. Prog. Phys. **66**, 789 (2003).
3. A. Labeyrie, S. G. Lipson, and P. Nisenson, *An introduction to astronomical stellar interferometry*, Cambridge U. (2006).
4. J. B. Le Bouquin, J. Berger, P. R. Labeye, E. Tatulli, F. Malbet, K. Rousselet, and P. Kern, SPIE **5491**, 1362 (2004).
5. C. A. Hummel, J. A. Benson, D. J. Huttter, K. J. Johnston, D. Mozurkewich, J. T. Armstrong, R. B. Hindley, G. C. Gilbreath, L. J. Pickard, and N. M. White, Astron. J. **125**, 2630 (2003).
6. G. Perrin, S. Lacour, J. Woillez, and E. Thiebaut, Mon. Not. Roy. Astr. Soc. **373**, 747 (2006).
7. J. B. Le Bouquin and E. Tatulli, Mon. Not. Roy. Astr. Soc. **372**, 639 (2006).
8. J. D. Armitage and A. Lohmann, Opt. Acta **12**, 185 (1965)
9. E. N. Ribak et al., Proc. ESO **29**, 1105, Garching, Germany (1988).
10. J. B. Breckinridge, E. Ribak, C. Roddier, F. Roddier and C. Habecker, *Real time optical correlation using white light Fourier transforms.* Final Report Tech. Work, Jet Propulsion Laboratory, Caltech, Pasadena, CA (1988).
11. E. N. Ribak, M. Gai, S. G. Lipson, and P. Parahovnik, SPIE **6268**-57 (2006).
12. I. Moreno, G. Paez, and M. Strojnik, Opt. Comm. **220**, 257 (2003).
13. E. J. Galvez and C. D. Holmes, J. Opt. Soc. Am. A **16**, 1981 (1999).